\documentstyle[12pt]{article}
\def\beq{\begin{equation}}
\def\eeq{\end{equation}}
\def\bea{\begin{eqnarray}}
\def\eea{\end{eqnarray}}
\def\nn{\nonumber}
\def\daga{A^{\dagger}}
\def\dagb{B^{\dagger}}
\def\qos{{\cal A}}
\def\qsos	{{\cal B}}
\def\super{$ SU_q(n/m) $ }
\begin{document}
\baselineskip = 18pt
\begin{tabbing}
\` RCNP-070 \\
\` March 1994 \\
\end{tabbing}
\begin{center}
{\Large Isomorphisms between Quantum Group Covariant
q-Oscillator Systems Defined for $ q $ and $ q^{-1} $}\\[2cm]
{\large N. Aizawa}\footnote{Address after April 1994,  Department
of Applied Mathematics, Osaka Women's University, Sakai, Osaka
590, Japan}\\[1cm]
{\em Research Center for Nuclear Physics}\\
{\em Osaka University, Ibaraki, Osaka 567, Japan}
\end{center}
\vfill
\begin{abstract}
It is shown that there exists an isomorphism between
q-oscillator systems covariant under $ SU_q(n) $ and $ SU_{q^{-1}}(n) $.
By the isomorphism, the defining relations of $ SU_{q^{-1}}(n) $
covariant q-oscillator system are transmuted into those of
$ SU_q(n) $. It is also shown that the similar isomorphism exists
for the system of q-oscillators covariant under the quantum supergroup
$ SU_q(n/m) $. Furthermore the cases of q-deformed Lie (super)algebras
constructed from covariant q-oscillator systems are considered.
The isomorphisms between q-deformed Lie (super)algebras can not
obtained by the direct generalization of the one for covariant
q-oscillator systems.\\
PACS 02.20.Tw, 02.20.Qs
\end{abstract}
\newpage

  Since the discovery of quantum deformation
(the so-called q-deformation)
of Lie groups and Lie algebras [1-5], many q-deformed objects have been
introduced. We can mention q-deformed hyperplane \cite{ma},
differential forms and derivatives on q-deformed hyperplane \cite{wz},
q-(super)oscillators \cite{har,ck},
q-deformed covariant oscillator systems [10-13]
and their generalization \cite{vdj}, q-symplecton \cite{nom1,nom2},
reflection equation algebras \cite{ks}, and so on.
Almost all of these objects are essentially defined by the same
algebraic structure, that is, Zamolodchikov-Faddeev algebra \cite{kul2}
or quantum group tensor \cite{kryu}.
However the relationship between q-deformed
objects defined for different values of the deformation parameter $q$
is unclear. This problem has been discussed for the q-oscillator
$ H_q = \{a, a^{\dagger}, N \} $ and found that
the central element of $ H_q $ plays a crucial
role. Assuming that the element $ N $ and the central element are independent
of $q$, Chaichian et al. derived the formula which transforms the elements
of $ H_{q_1} $ to the corresponding ones of $ H_{q_2} $ \cite{cpre}.
Without such assumption, the  present author found the one-to-one
correspondence between the elements of $ H_q $ and $ H_{q^{-1}} $ which
transmute the defining relations of $ H_{q^{-1}} $ into those of
$ H_q $ \cite{naru}. The elements of $ H_{q^{-1}} $ can be expressed
in terms of those of $ H_q $, therefore,
we can say that the algebra $ H_q $
is invariant under the replacement $ q \leftrightarrow q^{-1} $.
In mathematical language, $ H_q $ is isomorphic to $ H_{q^{-1}} $.

  In this article, it is shown that there exists an isomorphism between
q-oscillator systems which are covariant under $ SU_q(n) $ and
$ SU_{q^{-1}}(n) $.
By the isomorphism, the defining relations of $ SU_{q^{-1}}(n) $
covariant q-oscillator system are transmuted into those of
$ SU_q(n) $. It is also shown that the similar isomorphism exists
for the system of q-oscillators covariant under the quantum supergroup
$ SU_q(n/m) $. Furthermore q-deformed Lie (super)algebras constructed
from covariant q-oscillator systems are considered. They are also
covariant under the coaction of $ SU_q(n) $ and \super. It is shown
that, unfortunately, the isomorphisms between covariant q-oscillator
systems are not applicable to establish the isomorphisms
between q-deformed Lie (super)algebras.

  We start with the $ SU_q(n) $ covariant q-oscillator system
$ {\cal A}_q $. It is generated by $ 2n $ generators
$ \{ A_i, \daga_i,\ i = 1, \cdots n \} $ and they satisfy the following
defining relations \cite{pw,ckl,nom1}
\[
    A_i A_j = q A_j A_i, \qquad \daga_i \daga_j = q^{-1} \daga_j \daga_i,
     \qquad i < j
\]
\beq
    A_i \daga_j = q \daga_j A_i,                     \label{one}
\eeq
\[
    A_i \daga_i - q^2 \daga_i A_i
     = 1 + (q^2-1) \sum_{k =1}^{i-1}\; \daga_k A_k.
\]
Throughout this article, we assume to be $ q \in {\bf \rm R},\ q > 1 $.
The *-antiinvolution is introduced by
\beq
    (A_i)^* = \daga_i, \qquad (\daga_i)^* = A_i.             \label{two}
\eeq
The q-annihilation operators and the q-creation operators are covariant
and contravariant tensors of rank 1 under the coaction of $ SU_q(n) $,
respectively. It means that the transformations
\[
   A_i \ \ \rightarrow \ \ A_i' = \sum_{k=1}^n \; t_{ij}\; A_j,
\]
\beq
   \daga_i \ \ \rightarrow \ \
   \daga_i{}' = \sum_{k=1}^n \; t_{ij}^* \; \daga_j, \label{three}
\eeq
preserve the defining relations of $ {\cal A}_q $ (\ref{one}).
Here we denote
the generators of $ SU_q(n) $ by $ T = (t_{ij}) $ in matrix form and
assume that $ t_{ij} $ commute with all the generators of $ {\cal A}_q $.
The commutation relations of $ t_{ij} $ are written using the R-matrix
\beq
   R\; T \otimes T = T \otimes T \; R,    \label{four}
\eeq
the R-matrix is given by
\beq
  R = q\; \sum_{i=1}^n\; e_{ii} \otimes e_{ii}
    + \sum_{i \neq j}\; e_{ii} \otimes e_{jj}
    + (q - q^{-1})\; \sum_{i > j}\; e_{ij} \otimes e_{ji},  \label{five}
\eeq
where $ e_{ij} $ is the $ n \times n $ matrix with entry 1 at
position $ (i,j) $ and 0 elsewhere. The *-antiinvolution is defined by
\[
  T^* = (t_{ji}^*),\qquad T^* T = T T^* = I,
\]
where $ I $ is the $ n \times n $ unit matrix. The quantum determinant,
which is the central element of $ GL_q(n) $,
is defined by
$
 det_q T = \sum_{\sigma}\;(-q)^{l(\sigma)} t_{1\;\sigma(1)} \cdots
 t_{n\;\sigma(n)}
$,
where $ l(\sigma) $ is the minimal number of inversions in the
permutation $ \sigma $. We set $ det_q T = 1 $.

  We also define the $ SU_{q^{-1}} $ covariant q-oscillator system
$ \qos_{q^{-1}} $ by replacing $q$ with $q^{-1}$. in (\ref{one}). They
form rank 1 tensors under the coaction of $ SU_{q^{-1}} $. We denote
q-deformed objects defined for $q^{-1}$ by attaching bar on their elements,
{\em e.g.} $ \qos_{q^{-1}} = \{ \bar{A}_i, \bar{\daga}_i \} $. Our aim is
to establish a relationship between $ \qos_q $ and $ \qos_{q^{-1}} $.
It should be noted that the trivial relation
$ \bar{A}_i = A_i,\ \bar{\daga}_i = \daga_i $ is prevented, since
it concludes an unacceptable results ;
$ A_i A_j = 0,\ etc. $

  It is possible to relate the elements of $ \qos_q $ and
$ \qos_{q^{-1}} $ so that the covariant q-oscillator system is
invariant under $ q \leftrightarrow q^{-1} $. We prove the following
statement.

  There exists an isomorphism
$ \varphi : \qos_q \rightarrow \qos_{q^{-1}} $ such that $ \varphi $
transmutes the defining relations of $\qos_{q^{-1}}$ into those of
$ \qos_q $. The explicit formulae are given by
\beq
   \bar{A}_i = \Gamma_{i-1}^{-1} \Gamma_i^{-1} A_i, \qquad
   \bar{\daga}_i = \daga_i \Gamma_{i-1}^{-1} \Gamma_i^{-1} \label{six}
\eeq
where
\beq
   \Gamma_i \equiv \sqrt{[A_i, \daga_i]},\qquad \Gamma_0 \equiv 1.
                                                           \label{seven}
\eeq

\noindent
({\em Sketch of Proof})

  Using the properties of $\Gamma_i $, we can prove the statement
by direct calculations. Note that, using (\ref{one}),
$ \Gamma_i \ (i \neq 0) $ is rewritten as
\bea
  \Gamma_i & = \sqrt{A_{i+1} \daga_{i+1} - q^2 \daga_{i+1} A_{i+1}}
           \nn \\
           & = \sqrt{1 + (q^2-1) \displaystyle{
           \sum_{k=1}^{i} \daga_k A_k}}.
                                                           \label{eight}
\eea
$ \Gamma_i $ is not affected by the *-antiinvolution ;
$ \Gamma_i^* = \Gamma_i $. From these facts, we obtain the useful
relations
\bea
  & [\Gamma_i, \Gamma_j] = 0, \nn \\
  & A_i \Gamma_j = q \Gamma_j A_i, \qquad
    \daga_i \Gamma_j = q^{-1} \Gamma_j \daga_i, \qquad i \leq j
                                                         \label{nine} \\
  & [A_i, \Gamma_j] = [\daga_i, \Gamma_j] = 0. \qquad i > j \nn
\eea

  As an illustration, we take the last relation in (\ref{one})
\beq
  \bar{A}_i \bar{\daga}_i - q^{-2} \bar{\daga}_i \bar{A}_i
  = 1 + (q^{-2}-1) \sum_{k = 1}^{i-1}\; \bar{\daga}_k \bar{A}_k.
                                                           \label{ten}
\eeq
Substituting (\ref{six}) into (\ref{ten}) and multiplying
$ \Gamma_{i-1} \Gamma_i $ from both left and right, we obtain
\beq
   A_i \daga_i - \daga_i A_i
   =  \Gamma_{i-1}^2 \Gamma_i^2 \; \{1-(q^2-1)\;
        \sum_{k = 1}^{i-1}\; \Gamma_{k-1}^{-2} \Gamma_k^{-2}
        \daga_k A_k \}.
                                                        \label{eleven}
\eeq
Here the properties of $\Gamma_i$ (\ref{nine}) were used. Because of
the identity
\beq
  \Gamma_i^2  \{ 1 - (q^2-1)\; \sum_{k=1}^i\; \Gamma_{k-1}^{-2}
  \Gamma_k^{-2}\; \daga_k A_k \} = 1,                   \label{twelve}
\eeq
(\ref{eleven}) reads
\bea
  A_i \daga_i - \daga_i A_i
  & = & \Gamma_i^2 \nn \\
  & = & 1 + (q^2-1) \; \sum_{k=1}^i\; \daga_k A_k. \nn
\eea
Rearranging $ \daga_i A_i $, we obtain the last relation in (\ref{one}).

  The identity (\ref{twelve}) is proved by mathematical induction.
For $ i= 1 $, the left hand side of (\ref{twelve}) reads
\bea
  & & \Gamma_1^2 \{ 1 - (q^2 -1)\; \Gamma_1^{-2} \daga_1 A_1 \} \nn \\
  & = & \Gamma_1^2 \Gamma_1^{-2} \{ \Gamma_1^2 - (q^2-1) \daga_1 A_1 \} \nn \\
  & = & 1. \nn
\eea
Assuming that (\ref{twelve}) is valid for $ \Gamma_i $, consider
the case for $ \Gamma_{i+1} $
\bea
  & & \Gamma_{i+1}^2 \{ 1 - (q^2-1)\; \sum_{k=1}^{i+1}\;
      \Gamma_{k-1}^{-2} \Gamma_k^{-2} \daga_k A_k \}   \nn \\
  & = & \Gamma_{i+1}^2 \{ \Gamma_i^{-2} - (q^2-1)\;
      \Gamma_i^{-2} \Gamma_{i+1}^{-2} \daga_{i+1} A_{i+1} \} \nn \\
  & = & \Gamma_i^{-2} \{ \Gamma_{i+1}^2 - (q^2 -1)\;
      \daga_{i+1} A_{i+1} \} \nn \\
  & = & \Gamma_i^{-2} \Gamma_i^2 = 1. \nn
\eea
The identity (\ref{twelve}) has been proved.

  From (\ref{six}), it is obvious that $ \varphi $ is a one-to-one
correspondence. In order to show that $ \varphi $ is an isomorphism,
let us consider $ \varphi' : \qos_{q^{-1}} \rightarrow \qos_q $ defined
by
\[
  A_i = \bar{\Gamma}_{i-1}^{-1} \bar{\Gamma}_i^{-1} \bar{A}_i, \qquad
  \daga_i = \bar{\daga}_i \bar{\Gamma}_{i-1}^{-1} \bar{\Gamma}_i^{-1},
\]
and show that $ \varphi \circ \varphi' = \varphi' \circ \varphi = 1 $.
To this end, it is enough to note the relation
\bea
  \bar{\Gamma}_i & = & \sqrt{[ \bar{A}_i, \bar{\daga}_i ]} \nn \\
                 & = & \sqrt{\Gamma_{i-1}^{-2} \Gamma_i^{-2}
                       ( A_i \daga_i - q^2 \daga_i A_i) } \nn \\
                 & = & \Gamma_i^{-1}. \nn
\eea
Therefore the statement has been proved.

  It is emphasized that, in the limit of $ q \rightarrow 1 $, the
both hand sides of (\ref{six}) are reduced to the same bosonic
oscillators.

\bigskip

  Let us next consider the q-deformed Lie algebra $ u_q(n) $ which
is constructed from $ \qos_q $. As in the limit of
$ q \rightarrow 1 $, the bilinear forms of q-creation and
q-annihilation operators can define $ u_q(n) $,
\beq
  E_{ij} = \daga_i A_j, \qquad E_{ij}^* = E_{ji}  \label{thirteen}
\eeq
The commutation relations among $ E_{ij} $'s are obtained by using
(\ref{one}). In terms of the R-matrix, (\ref{one}) is rewritten as
follows,
\bea
  A_j A_i & = & q^{-1} \displaystyle{
            \sum_{kl}\; R_{ij,kl} A_k A_l}        \nn \\
  \daga_j \daga_i & = & q^{-1} \displaystyle{
            \sum_{kl}\; R_{kl,ij} \daga_k \daga_l} \\   \label{fourteen}
  A_j \daga_i & = & \delta_{ij} + q  \displaystyle{
            \sum_{kl}\; R_{il,kj} \daga_k A_l}  \nn
\eea
The commutation relations of $ u_q(n) $ are given by
\bea
   & & q \displaystyle{\sum_{abcdef}\; R^{t_1}_{ab,\mu\nu} R^{-1}_{a\rho, cd}
                         R_{cf, e\sigma} \; E_{ef} E_{db}} \nn \\
   & - & q^{-1} \displaystyle{ \sum_{abcdef}\; R_{\mu a, bc} R_{cd, e\nu}
                                R^{t_1}_{fe,\rho\sigma}\; E_{fd} E_{ba}}
                                                          \nn \\
   & = & q \displaystyle{ \sum_{ab}\; R^{t_1}_{ab,\mu\nu}
            \delta_{\rho\sigma} E_{ab}
     - q^{-1} \sum_{ab}\; R^{t_1}_{ab, \rho\sigma} \delta_{\mu\nu}
                          E_{ab}   }              \label{fifteen}    \\
   & - & q^{-1} \omega \displaystyle{\sum_{abc}\; R^{t_1}_{ab, \mu\nu}
                                    R^{t_1}_{ca, \rho\sigma} E_{cb} }\nn
\eea
where $ t_1 $ means the transposition in the first space and
$ \omega \equiv q - q^{-1} $.
This complicated relations are reduced to the usual commutation
relations of $ u(n) $ in the limit of
$ q \rightarrow 1 $
\[
  [E_{\mu\nu}, E_{\rho\sigma}] = \delta_{\nu\rho} E_{\mu\sigma}
  - \delta_{\mu\sigma} E_{\nu\rho},
\]
since the R-matrix is reduced to the unit matrix :
$ R_{ij,kl} \rightarrow \delta_{ik} \delta_{jl} $. From now on we
adopt the equation (\ref{fifteen}) as the defining relation of
$ u_q(n) $ without the aid of covariant q-oscillator realization
(\ref{thirteen}). The Hopf algebra structure for this $ u_q(n) $
is still an open problem.

  The algebra $ u_q(n) $ forms the $ SU_q(n) $ tensor of rank (1,1),
that is, the relation (\ref{fifteen}) is preserved by the
transformation
\beq
   E_{ij} \ \ \rightarrow \ \ E_{ij}' = \sum_{kl}\; t_{ik}^* t_{jl} E_{kl}.
                                                 \label{sixteen}
\eeq
This can be proved by direct calculation using the properties of the
R-matrix {\em i.e.} (\ref{four}), Yang-Baxter equation and
$ R_{ij,kl} = R_{lk,ji} $.

  We expect that $ u_{q^{-1}}(n) $ is isomorphic to $ u_q(n) $
and the isomorphism is given by
\bea
   \bar E_{ij} & = & q^3 \Gamma_{i-1}^{-1} \Gamma_i^{-1} \Gamma_{j-1}^{-1}
   \Gamma_j^{-1} E_{ij}, \qquad i < j  \nn \\
   \bar E_{ii} & = & q^2 \Gamma_{i-1}^{-2} \Gamma_i^{-2} E_{ii}
                                                     \label{seventeen}
\eea
where $ \bar E_{ij} $ denotes the element of $ u_{q^{-1}}(n) $.
However it does not hold except the case of $ n=2 $. In order to
show it, let us first consider the case of $ n=2 $. For $ n=2 $,
(\ref{fifteen}) is reduced to
\bea
 & & [E_{11}, E_{22}] = 0,     \nn \\
 & & E_{11} E_{12} - q^2 E_{12} E_{11} = E_{12}, \qquad
     [E_{22}, E_{12}] = -E_{12} - (q^2-1) E_{12} E_{11}, \\ \label{eighteen}
 & & q^2 E_{12} E_{21} - E_{21} E_{12} = (q^2-1) E_{11}^2 + E_{11} - E_{22},
     \nn
\eea
and $ \Gamma_i \ (i = 1,2) $ are restricted to the second expression
in eq.(\ref{eight})
\beq
 \Gamma_1 = \sqrt{1 + (q^2-1) E_{11}}, \qquad
 \Gamma_2 = \sqrt{1 + (q^2-1) (E_{11} + E_{22})}.      \label{nineteen}
\eeq
It is easy to see that $ E_{11} + E_{22} $ is a central
element of this algebra, therefore, $ \Gamma_2 $ is also a central
element. The non-trivial commutation relations are given by
\beq
  \Gamma_1 E_{12} = q E_{12} \Gamma_1, \qquad
  \Gamma_2 E_{21} = q^{-1} E_{21} \Gamma_1.            \label{twenty}
\eeq
It is proved by direct calculation that the isomorphism between $ u_q(2) $
and $ u_{q^{-1}}(2) $ is given by eq.(\ref{seventeen}). Here, we give only
one example, the last relation of (\ref{eighteen})
\beq
   q^{-2} \bar E_{12} \bar E_{21} - \bar E_{21} \bar E_{12}
   = (q^{-2} - 1) \bar E_{11}^2 + \bar E_{11} - \bar E_{22}. \label{twentyone}
\eeq
After substituting (\ref{seventeen}) into (\ref{twentyone}), we can
arrange $ \Gamma_i $ to the left of $ E_{kl} $ by making use of
(\ref{twenty})
\[
 \Gamma_1^{-2} \Gamma_2^{-2} (q^2 E_{12} E_{21} - E_{12} E_{21})
 = (1 - q^2) \Gamma_1^{-2} E_{11}^2 + E_{11} - \Gamma_2^{-2} E_{22},
\]
where we dropped the common factor $ \Gamma_1^{-2} $. Multiplying
$ \Gamma_1^2 \Gamma_2^2 $ from the left, we obtain the last
equation of (\ref{eighteen}). Furthermore because of (\ref{seventeen}),
\[
  \bar \Gamma_1 = \Gamma_1^{-1}, \qquad \bar \Gamma_2 = \Gamma_2^{-1},
\]
hold. Therefore the isomorphism between $ u_q(2) $ and $ u_{q^{-1}}(2) $
has been proved.

  On the other hand, for $ n \geq 3 $,
$ \displaystyle{\sum_{i=1}^n\; E_{ii}} $ is no longer a central
element, so that the commutation relation between $ \Gamma_i $ and
$ E_{kl} $ becomes quite complicated and the mechanism which makes
$ u_q(2) $ be isomorphic to $ u_{q^{-1}}(2) $, namely arranging
$ \Gamma_i $ to the left of $ E_{kl} $, does not work. Therefore
eq.(\ref{seventeen}) does not give the isomorphism between
$ u_q(n) $ and  $ u_{q^{-1}}(n) $ for $ n \geq 3 $.

\bigskip

  The isomorphisms discussed here can be generalized to the
\super covariant q-oscillator system $ \qsos_q $ which is generated by
$ 2n $ even generators $ \{ A_i, \daga_i, \ i = 1, 2, \cdots , n \} $
and $ 2m $ odd generators
$ \{ B_r, \dagb_r,\ r = 1, 2, \cdots, m \}. $ The *-antiinvolution
of a generator without dagger gives the corresponding one with
dagger and vice versa. The $ 2(n+m) $ generators satisfy the
following defining relations \cite{ckl}
\bea
  & & A_i A_j = q A_j A_i,  \qquad \qquad i < j  \nn \\
  & & A_i \daga_j = q \daga_j A_i, \qquad \qquad i \neq j \nn \\
  & & A_i \daga_i - q^2 \daga_i A_i = 1 + (q^2-1) \sum_{k=1}^{i-1}\;
      \daga_k A_k,                               \nn \\
  & & A_i B_r = q B_r A_i, \qquad
      A_i \dagb_r = q \dagb_r A_i,              \label{twentytwo} \\
  & & B_r B_s = -q B_s B_r, \qquad \qquad r < s  \nn \\
  & & B_r \dagb_s = -q \dagb_s B_r, \qquad \qquad r \neq s \nn \\
  & & B_r \dagb_r + \dagb_r B_r = 1 + (q^2-1) \sum_{k=1}^n \;
      \daga_k A_k + (q^2-1) \sum_{s=1}^{r-1}\; \dagb_s B_s, \nn \\
  & & B_r^2 = ( \dagb_r )^2 = 0,                 \nn
\eea
and their *-involution. The algebra $ \qsos_q $ forms a rank 1 tensor of
\super. We assume that even generators $ \qsos_q $ commute with all of
\super, odd generators of $ \qsos_q $ commute with even ones of \super,
while they anticommute with odd ones of \super. The coaction of
\super on $ \qsos_q $ is defined by
\bea
   \alpha_i \quad & \rightarrow & \quad
   \alpha_i' = \sum_{j=1}^{n+m}\; t_{ij} \alpha_j, \nn \\
   \alpha_i^{\dagger} \quad & \rightarrow & \quad
   \alpha_i^{\dagger}{}' = \sum_{j=1}^{n+m} \; (-)^{p(t_{ij})\; p(\alpha_j)}
                     t_{ij}^* \alpha_j^{\dagger},  \label{twentythree}
\eea
where $ t_{ij} \in SU_q(n/m) $ and we introduced the unified
notations for $ \qsos_q $
\[
  \alpha_i = A_i \quad {\rm for} \ \ 1 \leq i \leq n, \qquad
  \alpha_{i+n} = B_i \ \ (1 \leq i \leq m). \ \ etc.
\]
And $ p(a) $ denotes the parity of operator $ a $, namely
$ p(a) = 1 $ for odd $ a $, $ p(a) = 0 $ for even $ a $.

  The R-matrix for \super is given by \cite{ck}
\beq
   R = \sum_i^{n+m}\; q^{1-2p(i)} e_{ii} \otimes e_{ii}
     + \sum_{i \neq j}\; e_{ii} \otimes e_{jj}
     + w \sum_{i > j}\; e_{ij} \otimes e_{ji},     \label{twentyfour}
\eeq
where $ p(i) $ denotes the parity of $i$ th basis vector. We prove
the following relationship between $ \qsos_q $ and $ \qsos_{q^{-1}} $.

    There exists an isomorphism
$ \varphi : \qsos_q \rightarrow \qsos_{q^{-1}} $ such that $ \varphi $
transmutes the defining relations of $\qsos_{q^{-1}}$ into those of
$ \qsos_q $. The explicit formulae are given by
\bea
   \bar{A}_i & = & \Gamma_{i-1}^{-1} \Gamma_i^{-1} A_i, \qquad
   \bar{\daga}_i = \daga_i \Gamma_{i-1}^{-1} \Gamma_i^{-1} \nn \\
   \bar{B}_r & = & \Lambda_{r-1}^{-1} \Lambda_r^{-1} B_r, \qquad
   \bar{\dagb}_r = \dagb_r \Lambda_{r-1}^{-1} \Lambda_r^{-1},
                                                   \label{twentyfive}
\eea
where
\bea
   \Gamma_i & \equiv & \sqrt{[A_i, \daga_i]},\qquad \Gamma_0 \equiv 1,
                                                              \nn \\
   \Lambda_r & \equiv & \sqrt{B_r \dagb_r + q^2 \dagb_r B_r}, \qquad
   \Lambda_0 \equiv \Gamma_n.
                                                   \label{twentysix}
\eea

\noindent
({\em Sketch of Proof})

  As in the case of $ SU_q(n) $ covariant q-oscillator, the statement
can be proved by direct calculations using the commutation relations
among $ \Gamma_i $, $ \Lambda_r $ and the generators of $ \qsos_q $.
Because of the last relation of eq.(\ref{twentytwo}),
$ \Lambda_r \ (r \neq 0) $ can be rewritten as
\beq
  \Lambda_r = \{ 1 + (q^2 - 1) \sum_{k=1}^n \; \daga_k A_k +
                 (q^2 - 1) \sum_{s=1}^r \; \dagb_s B_s \}^{1/2}.
                                                  \label{twentyseven}
\eeq
The *-antiinvolution does not change $ \Lambda_r $ ;
$ \Lambda_r^* = \Lambda_r $. Following useful relations can be
shown easily
\bea
 & & [ \Gamma_i, \Lambda_r ] = [ \Lambda_r, \Lambda_s ] = 0,  \nn \\
 & & [ B_r, \Gamma_i ] = [ \dagb_r, \Gamma_i ] = 0,           \nn \\
 & & A_i \Lambda_r = q \Lambda_r A_i, \qquad
     \daga_i \Lambda_r = q^{-1} \Lambda_r \daga_i,   \label{twentyeight} \\
 & & [ B_r, \Lambda_s ] = [ \dagb_r, \Lambda_s ] = 0,   \quad
     {\rm for } \ \ s < r     \nn \\
 & & B_r \Lambda_s = q \Lambda_s B_r, \qquad
     \dagb_r \Lambda_s = q^{-1} \Lambda_s \dagb_r,  \quad
     {\rm for } \ \ s \geq r     \nn
\eea
It is not difficult to prove the statement using these relations together
with eq.(\ref{nine}). As a showcase, we consider the last relation of
eq.(\ref{twentytwo}). Again, we denote the generators of
$ \qsos_{q^{-1}} $ by the operators with bar.
\beq
  \bar B_r \bar \dagb_r + \bar \dagb_r \bar B_r
  = 1 + (q^{-2}-1) \sum_{k=1}^n \; \bar \daga_k \bar A_k
      + (q^{-2}-1) \sum_{s=1}^{r-1} \; \bar \dagb_s \bar B_s.
                                                     \label{twentynine}
\eeq
We substitute (\ref{twentyfive}) into (\ref{twentynine}), then
using (\ref{nine}) and (\ref{twentyeight}) we obtain
\bea
  & &   B_r \dagb_r + q^2 \dagb_r B_r                            \nn \\
  & = & \Lambda_{r-1}^2 \Lambda_r^2 \{ 1 - (q^2-1) \sum_{k=1}^n \;
        \Gamma_{k-1}^{-2} \Gamma_k^{-2} \daga_k A_k
      - (q^2-1) \sum_{s=1}^{r-1} \; \Lambda_{s-1}^{-2} \Lambda_s^{-2}
        \dagb_s B_s \}                                            \nn \\
  & = &  \Lambda_{r-1}^2 \Lambda_r^2 \{ \Gamma_n^{-2}
      - (q^2-1) \sum_{s=1}^{r-1} \; \Lambda_{s-1}^{-2} \Lambda_s^{-2}
      \dagb_s B_s \}.                             \label{thirty}
\eea
The relation (\ref{twelve}) was used to derive the last line. As is
shown later, an analogous identity to (\ref{twelve}) holds
\beq
  \Lambda_r^2 \{\Gamma_n^{-2} - (q^2-1) \sum_{s=1}^r \;
  \Lambda_{s-1}^{-2} \Lambda_s^{-2} \dagb_s B_s \} = 1.  \label{thirtyone}
\eeq
Because of this identity, (\ref{thirty}) can be rewritten
\bea
   &  & B_r \dagb_r + q^2 \dagb_r B_r                         \nn \\
   & = & \Lambda_r^2                                          \nn \\
   & = & 1 + (q^2-1) \sum_{k=1}^n \; \daga_k A_k
           + (q^2-1) \sum_{s=1}^r \; \dagb_s B_s.             \nn
\eea
Rearranging $ \dagb_r B_r $, we obtain the last relation of
(\ref{twentytwo}).

  The identity (\ref{thirtyone}) is proved by mathematical
induction. For $ r = 1 $, the left hand side of
(\ref{thirtyone}) reads
\bea
 & & \Lambda_1^2 \{ \Gamma_n^{-2} - (q^2-1) \Lambda_1^{-2}
     \Lambda_0^{-2} \dagb_1 B_1 \}                             \nn \\
 & = & \Gamma_n^{-2} \{ \Lambda_1^2 - (q^2-1) \dagb_1 B_1 \}   \nn
\eea
By definition of $ \Lambda_1 $, it is obviously reduced to unity.
Assuming that (\ref{thirtyone}) is valid for $ \Lambda_r $,
consider the case of $ \Lambda_{r+1} $
\bea
   & & \Lambda_{r+1}^2 \{ \Gamma_n^{-2} - (q^2-1) \sum_{s=1}^{r+1} \;
       \Lambda_{s-1}^{-2} \Lambda_s^{-2} \dagb_s B_s \}        \nn \\
   & = & \Lambda_{r+1}^2 \{ \Lambda_r^{-2} - (q^2-1) \Lambda_r^{-2}
         \Lambda_{r+1}^{-2} \dagb_{r+1} B_{r+1} \}             \nn \\
   & = & \Lambda_r^{-2} \{ \Lambda_{r+1}^2 - (q^2-1) \dagb_{r+1}
         B_{r+1} \}                                            \nn \\
   & = & 1.                                                    \nn
\eea
Therefore the identity (\ref{thirtyone}) has been proved.

  It can be easily seen that the map
$ \varphi' : \qsos_{q^{-1}} \rightarrow \qsos_q $ defined by
\bea
   A_i & = & \bar \Gamma_{i-1}^{-1} \bar \Gamma_i^{-1} \bar A_i, \qquad
   \daga_i = \bar \daga_i \bar \Gamma_{i-1}^{-1} \bar \Gamma_i^{-1} \nn \\
   B_r & = & \bar \Lambda_{r-1}^{-1} \bar \Lambda_r^{-1} \bar B_r, \qquad
   \dagb_r = \bar \dagb_r \bar \Lambda_{r-1}^{-1} \bar \Lambda_r^{-1},
                                                   \label{thirtytwo}
\eea
is the inverse map of $ \varphi $, because of the relation
\beq
   \bar \Lambda_r = \Lambda_r^{-1}.                \label{thirtythree}
\eeq
This completes the proof of the statement.

  It is natural to extend the isomorphism to the q-deformed Lie
superalgebra $ u_q(n/m) $ constructed from $ \qsos_q $. As in the
case of $ u_q(n) $, trivial extension is not valid against our
expectation. Let us show it in the simplest case $ u_q(1/1) $.
\cite{ckl} The generators of $ u_q(1/1) $ are constructed by
\beq
  Q = \daga B, \quad Q^{\dagger} = \dagb A, \quad
  X = \daga A, \quad Y = \dagb B.              \label{thirtyfour}
\eeq
They satisfy the following commutation relations
\bea
  & & Q^2 = 0,                                             \nn \\
  & & q^2 Q Q^{\dagger} + q^{-2} Q^{\dagger} Q
      = X + q^{-2} Y + (q^2-1) X^2,                        \nn \\
  & & X Q - q^2 Q X = Q,                      \label{thirtyfive}\\
  & & Y Q = 0, \qquad  q^2 Q Y = Q + (q^2-1) X Q,          \nn \\
  & & [X, Y] = 0,                                          \nn
\eea
and their *-involution. We regard (\ref{thirtyfive}) as the
defining relations of $ u_q(1/1) $ without the aid of
(\ref{thirtyfour}). According to (\ref{thirtyfour}), we
expect that $ u_{q^{-1}} $ is isomorphic to $ u_q(1/1) $ and the
isomorphism is given by
\bea
  & & \bar Q = q^3 \Gamma^{-2} \Lambda^{-1} Q, \qquad
      \bar Q^{\dagger} = q^3 Q^{\dagger} \Gamma^{-2} \Lambda^{-1},
                                                           \nn \\
  & & \bar X = q^2 \Gamma^{-2} X, \qquad
      \bar Y = q^2 \Gamma^{-2} \Lambda^{-2} Y,   \label{thirtysix}
\eea
where
\beq
  \Gamma = \sqrt{1 + (q^2-1) X}, \qquad
  \Lambda = \sqrt{1 + (q^2-1) (X+Y)}.            \label{thityseven}
\eeq

  As an example, we consider the second equation of (\ref{thirtyfive})
\[
  q^{-2} \bar Q \bar Q^{\dagger} + q^2 \bar Q^{\dagger} \bar Q
  = \bar X + q^2 \bar Y + (q^{-2}-1) \bar X^2.
\]
Substituting (\ref{thirtysix}) into this equation and
multiplying $ q^{-2} \Gamma^2 \Lambda^2 $ from the left,
we obtain
\[
  q^2 ( Q Q^{\dagger} + Q^{\dagger} Q )  = X + (q^2-1) X^2
      + q^2 Y + (q^4-1) X Y.
\]
The correct equation can not be derived unless the relation
\beq
  q^2 X Y + Y - Q^{\dagger} Q = 0,                \label{thirtyeight}
\eeq
holds. However eq.(\ref{thirtyeight}) does not hold
without the aid of the covariant q-oscillator realization
(\ref{thirtyfour}). Therefore we have shown that
(\ref{thirtysix}) does not give the isomorphism between
$ u_q(1/1) $ and $ u_{q^{-1}}(1/1) $.

\bigskip

  In this article, we have shown that, in the case of $ SU_q(n) $
and \super, the covariant q-oscillator systems defined for $ q $ are
isomorphic to the ones for $ q^{-1} $. The final goal of
investigation along the line presented here is to establish
relationships between all kinds of q-deformed objects defined for
$ q $ and $ q^{-1} $. This is not a easy but a challenging
problem. As has seen in the case of q-deformed Lie algebra,
the established isomorphism between covariant q-oscillators
can not be generalized directly to other q-deformed objects.
For the q-deformed Lie algebras, we will have to reanalyze the
isomorphism based on the structure of the algebra itself
without the aid of covariant q-oscillator realizations. We
can mention an another example, namely extended covariant
q-oscillator system. It is an algebra which consist of
some copies of a covariant q-oscillator system. The mutual
relationships among various copies should also be covariant
under the quantum group coaction. This requirement concludes
that generators of a covariant q-oscillator system do not
commute with their copies, and the commutation relations
among various copies becomes non-trivial. For example,
commutation relations between $ SU_q(n) $ covariant
q-oscillator system $ \qos_q = \{ A_i, \ \daga_i \} $
and its copy $ \{ D_i, D^{\dagger}_i \} $ are given,
in terms of the R-matrix, by \cite{nom1}
\bea
 & & D_j A_i = \displaystyle{q \sum_{kl}\; R_{ij, kl} A_k D_l},
     \qquad
     \daga_j D^{\dagger}_i = \displaystyle{q \sum_{kl}\; R_{kl, ij}
     D^{\dagger}_k \daga_l},                               \nn \\
 & & A_j D^{\dagger}_i = \displaystyle{q \sum_{kl}\; R_{il, kj}
     D^{\dagger}_k A_l},
     \qquad
     D_j \daga_i = \displaystyle{q \sum_{kl}\; R_{il, kj}
     \daga_k D_l}.                                        \label{thirtynine}
\eea
It can be easily verified that (\ref{six}) does not give the
isomorphism between this extended $ \qos_q $ and the one defined for
$ q^{-1} $, although we do not give the proof here. This is due to
the additional structure given by (\ref{thirtynine}), we have to
take it into consideration if we wish to establish the isomorphism.

  One of the most important problem concerning the isomorphism
discussed here is the relationships between quantum groups defined
for $ q $ and $ q^{-1} $, {\em e.g.} $ SU_q(n) $ and $ SU_{q^{-1}}(n) $.
Unfortunately, the result of the present article seems not to be
applicable to the problem. It will be a future work.

%

\end{document}